\documentclass{aanew}
\usepackage{graphicx}

\begin{document}

\title{A hidden population of Wolf-Rayet stars in the massive galactic cluster
\object{Westerlund 1}. 
\thanks{Based on observations collected at the European Southern
Observatory, La Silla, Chile (ESO 67.D-0211)}} 
\author{J.~S.~Clark\inst{1} 
\and I. Negueruela\inst{2,3} 
}

\offprints{J. S. Clark, \email{jsc@star.ucl.ac.uk}}

\institute{ Department of Physics and Astronomy, University College London,
Gower Street, London, WC1E 6BT, England, UK
\and
Observatoire de Strasbourg, 11 rue de l'Universit\'{e},
F67000 Strasbourg, France
\and
Dpto. de F\'{\i}sica, Ingenier\'{\i}a de Sistemas y Teor\'{\i}a de
la Se\~{n}al, Universidad de Alicante, Apdo. 99, E03080 Alicante, Spain}

\date{Received    / Accepted     }

\abstract{We report the discovery of a hitherto undetected population
of Wolf-Rayet stars 
in the young galactic open cluster \object{Westerlund~1}. Optical spectroscopy of the 
cluster  identified 11 such objects; provisional classification suggests that  6 are nitrogen rich 
(WN) and 5 carbon rich (WC). Including the previously identified Blue, Yellow  \& Red Super- \& Hypergiants,
\object{Westerlund~1} clearly has a very rich population of massive post-Main Sequence objects. 
To date, the post-MS  population of \object{Westerlund~1}   is significantly larger than that of any 
other galactic young   open  cluster - with the possible exception of the \object{Arches} -
implying that it is potentially amongst the most massive young clusters yet identified in the Local Group.
}

\titlerunning{Wolf-Rayet stars in Wd~1}

\maketitle

\section{Introduction}   

The highly reddened young open cluster \object{Westerlund~1}
(henceforth Wd~1) was first  
identified by Westerlund (\cite{westerlund61}).
Subsequent broadband photometric surveys by Borgman et
al. (\cite{borgman}), Lockwood  ({\cite{lockwood}) and Koornneef
(\cite{koornneef}) suggested the presence of a number of both early  
and late type supergiants, while a comprehensive photometric and
spectroscopic survey of the  
brightest cluster members was presented by Westerlund
(\cite{westerlund87}; West87). Despite reporting the 
presence   of  a large number of very luminous
($L >$10$^5$~$L_{\odot}$) transitional objects   
only one further (photometric) study of the cluster has been
made (Piatti et al. \cite{piatti}).  

Recently,  radio continuum observations of \object{Wd~1} revealed that
a  number of the cluster  members appeared to be associated with very
bright radio sources (Clark et al. \cite{clark98},  
Dougherty, Clark \& Waters, in prep.). Motivated by these results we obtained 
low resolution optical spectroscopy of a number of 
the brighter cluster members in order to provide an accurate 
spectral classification for them. In this paper we present the first results of 
this  program; the discovery of a significant population of Wolf-Rayet
(WR) stars within the cluster.

\section{Observations}
Spectroscopy of cluster members over the red/near-IR spectral region
($\sim$6000-11000{\AA}) was taken on 2001 June 23--25 from 
the ESO 1.52-m telescope at La Silla Observatory, Chile. The telescope
 was equiped with  the Loral \#38 camera and the \#1 (night 1) and  
\#13 (night 2 and  3) gratings, giving dispersions of $\sim 5$ \AA/pixel 
and  $\sim 7$ \AA/pixel - leading to resolutions of $\approx 11$\AA\ and   
$\approx 16$\AA\ - respectively. Data reduction was accomplished with 
packages within the {\em Starlink} software suite.

Due to the crowded nature of the field, each long slit integration 
included a number of different cluster members. 
Examination of the fainter objects present in several of the exposures 
revealed the presence of a number of objects 
with  rich emission line spectra (see Figs. 1 and  2). Given that the 
integrations were optimised to avoid saturating on the brighter 
cluster members, the serendipitous sources are of a low S/N ratio - though 
sufficient to identify the emission line objects as a previously 
unidentified population of massive, hydrogen depleted WRs.

\section{Results}

Despite the low S/N of many of the spectra, it is immediately possible 
to identify both nitrogen rich WN (6 objects) and
carbon rich WC (5 objects) stars (Figs. 1 and 2 respectively); 
a finding chart and co-ordinates for each object
are presented in Fig.~3 and Table~1.  

Accurate determination of the spectral types of the WN and WC 
stars in this spectral region is difficult, given that most commonly used 
diagnostics lie at shorter wavelengths. However initial spectral classification of 
the WR candidates using the catalogues of Vreux et al. (\cite{vreux83}, \cite{vreux90})
was possible. 

For preliminary classification of the WC candidates we  use the ratio of 
the C\,{\sc iii}(8500~{\AA})/C\,{\sc iv}(8856~{\AA}) and 
C\,{\sc ii}(9900~{\AA})/C\,{\sc iii}(9710~{\AA}) 
lines (Vreux et al. \cite{vreux83}; Howarth \& Schmutz \cite{howarth};
Crowther priv. comm.).   
C\,{\sc iv}(8856~{\AA}) is absent from the spectra of all candidates, 
with an upper limit to emission in candidates F, E and C constraining 
spectral types to later than WC6. No constraints are possible for H and K 
due to weak or no emission in C\,{\sc iii}(8500~{\AA}). 
C\,{\sc ii}(9900~{\AA})/C\,{\sc iii}(9710~{\AA}) ratios of 0.17$\pm$0.04, 
0.18$\pm$0.03 and 0.18$\pm$0.04 for candidates F, E  and H respectively, 
suggest WC9 classifications while a 
ratio of $<$0.08 suggests a WC8 classification for 
candidate C\footnote{EW(CII)/EW(CIII)=0.05-0.08 for WC8 
(from \object{WR~135, 113 and 53}) and EW(CII)/EW(CIII)=0.14-0.3 for WC9 
(from \object{WR~92 and 104})}, while 
a classification for candidate K is not possible.  

Classification of the WN spectra proved more difficult, as there are fewer lines present and there 
is no linear progression in e.g. the  N\,{\sc iv}:He\,{\sc ii} ratios that might be used for classification
(Vreux et al. \cite{vreux83}). 
Based on the strength of the N\,{\sc iv} 7103-7128{\AA} feature we  can exclude extreme WN3/WN9
subtypes for all objects, since these are the only spectral types for which it is not  seen in emission. 
Candidate A appears  to be a WNE (WN4-5) given the  large lines widths ($\sim$3000~kms$^{-1}$)
and lack of  
strong He\,{\sc i} 7065{\AA} emission (seen for all stars between WN6-8; Vreux et al. \cite{vreux83}, \cite{vreux90}). 
On the basis of emission  in this line and the small emission line widths, candidates G, I and D appear to be WN6-8 objects. 
Classification of candidates B and J is complicated by the low S/N in the blue 
regions of the spectra due to the CCD response curve, but are also  probably WNL objects.

\begin{figure}
\resizebox{\hsize}{!}{\includegraphics[angle=0]{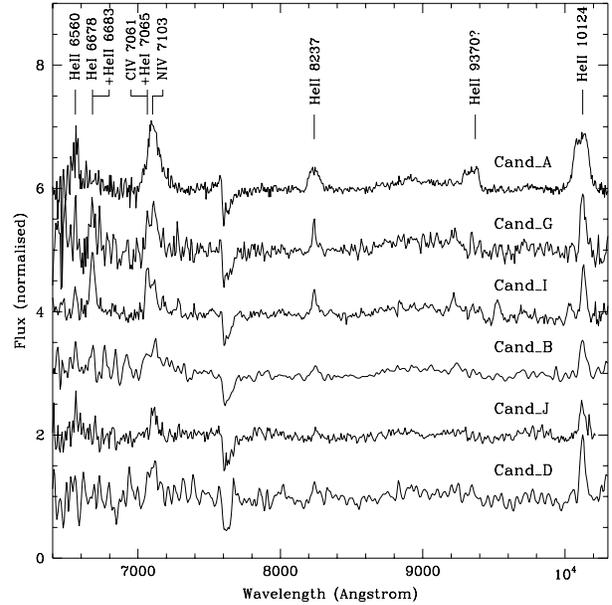}}
\label{Figure1}
\caption{Spectra of  the newly discovered WN candidates in \object{Wd~1}, 
with prominent transitions identified.}
\end{figure}

\begin{figure}
\resizebox{\hsize}{!}{\includegraphics[angle=0]{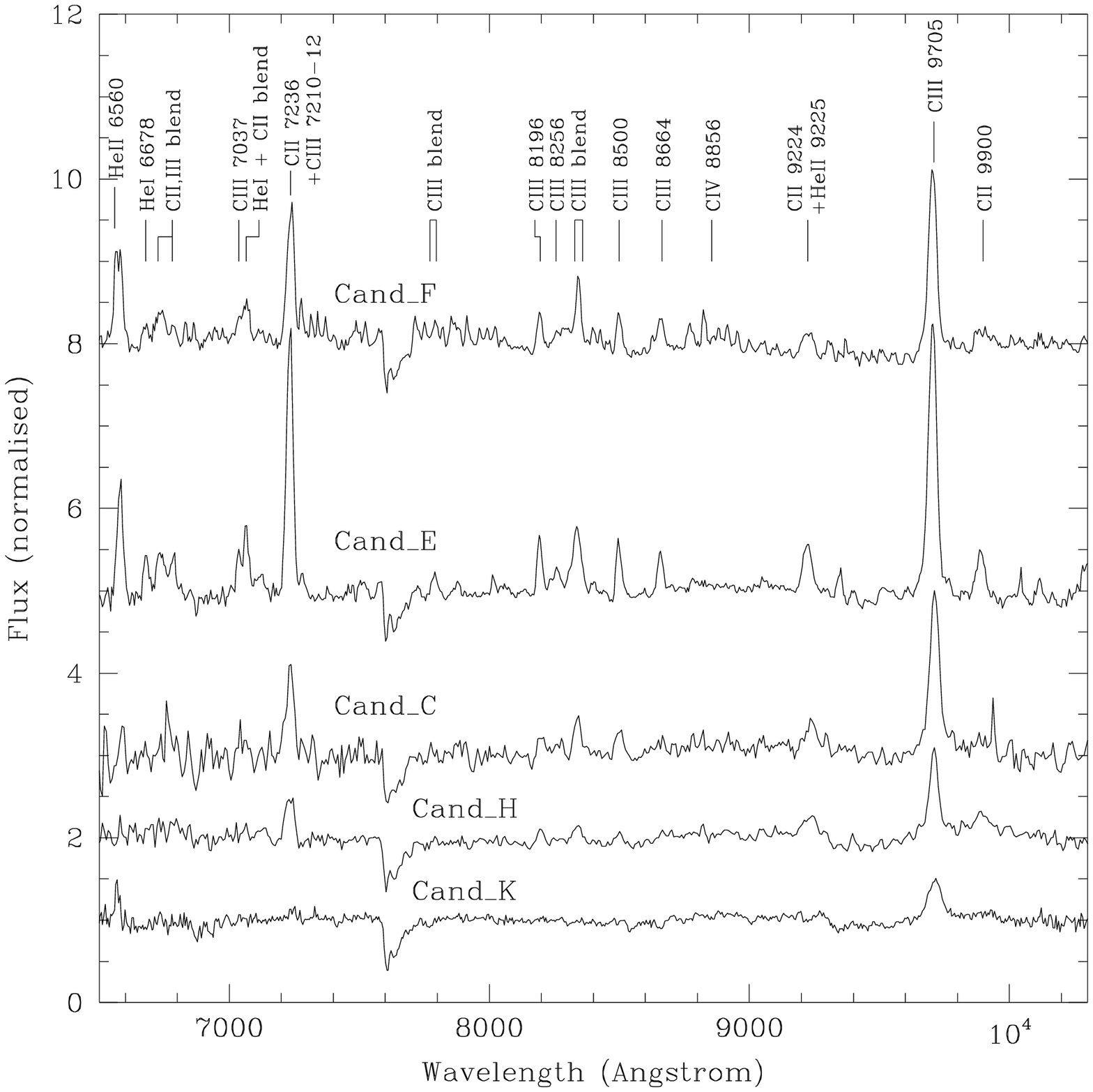}}
\caption{Spectra of  the newly discovered WC candidates in \object{Wd~1} with prominent transitions identified.}
\label{Figure2}
\end{figure}

\begin{figure*}
\resizebox{\hsize}{!}{\includegraphics[angle=-90]{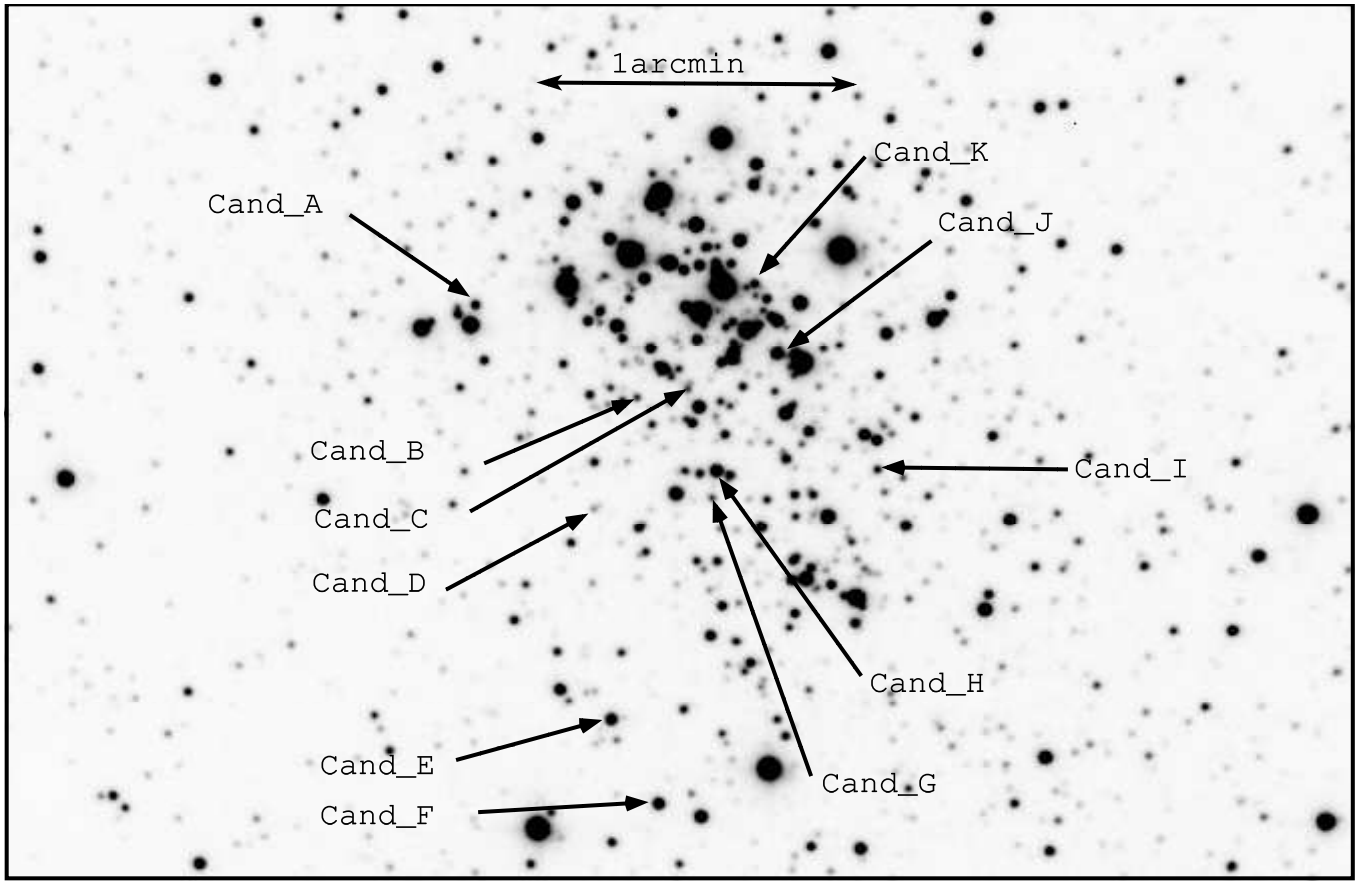}}\label{Figure 1}
\label{Figure3}
\caption{R broadband finding chart for the newly discovered Wolf Rayet candidates in \object{Wd~1}. 
Note that the exact counterparts of  candidates H, J \& K  should  be regarded as provisional 
since the crowded  field in these  regions of the cluster make identification of the correct counterpart difficult.
 Candidates A, E \& F correspond to stars number 72, 241 \& 239 respectively in the 
notation of West87.}
\end{figure*}  

\begin{table}
\caption{Co-ordinates (J2000) for the newly identified Wolf Rayet stars
in \object{Wd~1} determined  from 3.6cm radio images (Dougherty, Clark \& Waters; in prep.)
Formal errors are  $\sigma_\alpha = \pm0.003^{\rm s}$, and $\sigma_\delta= \pm0.04^{\prime\prime}$, although given that the crowded fields
in the vicinities of candidates H, J \& K make identification of the correct counterpart difficult, the errors for these objects are 
likely to  exceed the formal error quoted.}
\begin{center}
\begin{tabular}{lcc}
\hline
Candidate & $\alpha$ & $\delta$ \\
\hline
A & 16h47m8.324s & $-$45$^{o}$50$^{\prime}$45.51$^{\prime\prime}$\\
B & 16h47m5.354s & $-$45$^{o}$51$^{\prime}$05.03$^{\prime\prime}$\\
C & 16h47m4.395s & $-$45$^{o}$51$^{\prime}$03.79$^{\prime\prime}$\\
D & 16h47m6.243s & $-$45$^{o}$51$^{\prime}$26.48$^{\prime\prime}$\\
E & 16h47m6.056s & $-$45$^{o}$52$^{\prime}$08.26$^{\prime\prime}$\\
F & 16h47m5.213s & $-$45$^{o}$52$^{\prime}$24.97$^{\prime\prime}$\\
G & 16h47m4.015s & $-$45$^{o}$51$^{\prime}$25.15$^{\prime\prime}$\\
H & 16h47m3.905s & $-$45$^{o}$51$^{\prime}$19.88$^{\prime\prime}$\\
I & 16h47m1.668s & $-$45$^{o}$51$^{\prime}$20.40$^{\prime\prime}$\\
J & 16h47m0.885s & $-$45$^{o}$51$^{\prime}$20.85$^{\prime\prime}$\\
K & 16h47m2.697s & $-$45$^{o}$50$^{\prime}$57.35$^{\prime\prime}$\\
\hline
\end{tabular}
\end{center}
\end{table}

\section{Discusion \& Conclusions}

Unfortunately, uncertainty in the distance and reddening estimates to
\object{Wd~1} (West87; Piatti et al. \cite{piatti}),  
coupled with the poorly determined  bolometric corrections for many massive transitional objects make determination of the 
luminosities of the WRs and the other evolved cluster members difficult.
Indeed,  the lack of  accurate luminosity estimates for the  yellow hypergiant candidates (YHG; West87)
is particularly concerning,  given that  bolometric luminosity is one of several classification 
criteria for such objects (e.g., de Jager et al. \cite{dejager}). If
we are to constrain the post-MS population of  
\object{Wd~1} in order to determine such fundamental properties as
cluster age and mass we must address these issues.  

For the WR  candidates A, E and F (for which West87 provide photometry), adopting the 
reddening  and distance estimates given by West87\footnote{note that 
radio observations of the WR stars suggest that the cluster is more distant than 
the alternative distance of 1~kpc proposed by Piatti et al. \cite{piatti}; see  
Dougherty et al., in prep.} and the (conservative) bolometric corrections of 
Smith et al. (\cite{smith}) we find that their luminosities comfortably exceed 
10$^5 L_{\odot}$. For luminous A, F and  G stars  
the strength of the O\,{\sc i} 7774{\AA} feature can be used to
provide an additional measure of the absolute  
visual magnitude\footnote{Note
that our values of  EW(O\,{\sc i})  are consistent with  
those given by West87; Negueruela \& Clark (in prep).}.
Comparison of the YHG candidates W4, W12, W16 and 
W265  and the YSG W7 
(notation from West87) to the  luminosity:line strength calibrations
of Ferro \& Mendoza  (\cite{ferro}) and Slowik \&  Peterson
(\cite{slowik}) indicates that they are all likely to be
intrinsically highly luminous. Indeed the EW(O\,{\sc i}) 
for these objects are {\em significantly  greater than any} of the
stars in these studies -  including the  
the {\em bona fide} YHG \object{$\rho$ Cas}
($L_{\ast}$=10$^{5.7}L_{\odot}$) - 
implying luminosities of $L_{\ast} \gg $ 10$^{5}L_{\odot}$. Additionally, high mass loss
rates for these objects are implied by radio detections (W4, W12, W16 and W265) and broad H$\alpha$
emission   (W7, W12, W16 and W265); therefore all these objects meet the classification 
criteria of de Jager (\cite{dejager}) for {\em bona fide} YHGs.

\subsection{Comparison to other clusters}

Including the WRs and YHG candidates, the  large population of luminous  post-MS 
objects in \object{Wd~1} (West87 and  Table 4) 
suggests that it is unique in both the number and variety of massive 
post-MS  objects present. Of the 26 clusters within the solar circle studied
by Eggenberger et al. (\cite{eggenberger}), 6 contain both  B and RSGs.
  Of these, only  3 - \object{Collinder~228}, \object{Trumpler~27} and \object{Berkley~87} - 
also contain WRs (1-2 per cluster; Table 4) and only  
\object{Trumpler~27} contains a yellow SG, albeit of significantly lower luminosity ($\sim$10$^{4.7}L_{\odot}$) 
than those in \object{Wd~1} (Massey et al. \cite{massey01};
note however that they claim the cluster is {\em not} co-eval.).

Of the Galactic Centre clusters, the \object{Quintuplet cluster}
appears to contain 8  
WRs of both WN and WC types along with a number of early OB supergiants 
and a single RSG (10$^{4.9}L_{\odot}$; Figer et al. \cite{figer}). 
In addition to the single luminous RSG \object{IRS~7} (10$^{5.4}L_{\odot}$; Carr et al. 
\cite{carr}),  Paumard et al. (\cite{paumard}) detect 16 ``helium stars''  
in the \object{Galactic Centre cluster}, of which they suggest the 7
narrow line objects correspond to   mass losing BSGs (possible LBVs) and the 9 broad line
 objects WRs (noting that a further 3 stars  Blum et al. (\cite{blum96})  suggest are WCs lie outside their f-o-v). 
Only the \object{Arches cluster} - which is significantly younger - appears to contain a comparable number of 
WRs to \object{Wd~1}, with  Blum et al. (\cite{blum01}) identifying 15 candidate O4~If/WN7 stars  
on the basis of narrow band imaging. However, as with the \object{Quintuplet} and \object{Galactic Centre} clusters, the 
rich population of very luminous cool stars present in \object{Wd~1} is absent \footnote{the  very young clusters 
\object{NGC~3603} and \object{R~136} also contain a few high luminosity  WRs although these are  thought to be 
very massive O stars where the high mass loss rates simulate the spectra of more
chemically evolved lower mass WRs (e.g. de Koter et al. \cite{dekoter}).}. 

Clearly, such comparisons indicate that \object{Wd~1} is both very  young and very massive;
however uncertainties in the post-MS evolution of massive stars, exacerbated
by the lack of an identifiable MS turnoff and accurate bolometric luminosities for the evolved stars
makes  determination of the age and total mass of \object{Wd~1} difficult.

\subsection{The age and  mass of \object{Wd~1}}

Following the analysis of the \object{Quintuplet cluster} by Figer et al. (\cite{figer}),
the presence of WC  stars - apparently the most evolved stars present in \object{Wd~1} - 
implies a lower limit to the age of 2.5~Myr.
The presence of a number of very luminous RSGs within \object{Wd~1} potentially provides 
an upper limit to the cluster age; Figer et al. (\cite{figer}) suggests that the \object{Quintuplet 
cluster} requires an age of $\geq$4~Myr given the presence of a single (low luminosity) RSG,
 broadly consistent with the age (7~Myr) Carr et al. (\cite{carr}) 
claim for  \object{IRS~7}. However, large 
uncertainties in the mass loss rate for very luminous cool stars render estimates of their ages, lifetimes
 and progenitor masses highly uncertain; particularly concerning given the large population of YHGs 
 within \object{Wd~1}. 

Considering their  extreme rarity, the lifetime of YHGs is probably less than 10$^5$~yr for any 
luminosity and progenitor mass. At very high  luminosities it is  likely  that a very large  mass 
loss rate limits the YHG to a single passage from red to blue across the HR diagram, resulting in 
a  short lifetime ($\leq$30,000~yr; Stothers \& Chin \cite{stothers99}). At lower luminosities 
dynamical instabilites in the outer atmosphere of the star result in multiple blue loops for the 
star out of the RSG region, leading to a longer lifetime as a luminous yellow star (e.g. Stothers  
\& Chin \cite{stothers01}). Such estimates  are {\em qualitatively} consistent with the results of 
unpublished simulations for the behaviour of M$_{\rm initial}$=25 \& 40~$M_{\odot}$  stars (Maeder 
\&  Nieuwenhuijzen priv. comm. 2002) which suggest YHG phases after 6.9 \& 4.4~Myr  lasting  
$\sim$49000 and 2700~yr respectively; i.e. the YHG phase occurs earlier and 
is shorter   the more massive the progenitor is.

Despite the many uncertainties, the present stellar population of \object{Wd~1} appears consistent with an 
age  of order 4-8~Myr {\em if the cluster is co-eval}, suggesting it is potentially 
younger than previously thought (7 and 8$\pm$3~Myr; West87 \& Piatti et al. \cite{piatti}, respectively).   

The uncertainties in post-MS evolution and  incomplete stellar census inevitably makes a 
determination of the total cluster mass uncertain. Maeder \& Meynet (\cite{maeder}) find that for 
solar metallicities, adopting {\em twice} the  standard mass loss rate  results in the appearance of
 a WN phase for stars $\geq$25~$M_{\odot}$ -  consistent with the findings of  Massey et al.
 (\cite{massey01}) - and a WC phase at $\geq$40~$M_{\odot}$. Given that the super- and
hyper-giants are  less chemically  evolved than the WRs,  their progenitors are likely to
 have been less massive.

Regarding the completeness of our sample we  estimate that the spectral survey of potential cluster members of a similar visual 
magnitude as the fainter WR candidates is at best 33~per cent complete, while preliminary analysis of our low resolution optical
 spectroscopy (Negueruela \& Clark, in prep.) finds many additional
supergiant candidates, 
e.g.,  W70 and W71 (BSG), W32 and W33 (YSG) and W75 and W237
(RSG). Additionally, we might expect that if originally present  
very massive stars will have  been lost to SN; Maeder \& Meynet
(\cite{maeder}) suggest  that stars of $\geq$85~$M_{\odot}$ 
 will have a lifetime comparable to the lower estimate of the cluster age.

Nevertheless, {\em conservatively} assuming that the present stellar census 
is complete (see Table 4) and  furthermore that the progenitor masses for these objects were 
$\geq$30$M_{\odot}$ we can derive a lower limit to the {\em initial} cluster mass. 
Adopting a Salpeter mass function  ($N(M) \propto M^{-{\alpha}}$) with upper and lower cut 
offs of 100~$M_{\odot}$ \&  0.2~$M_{\odot}$ respectively and a slope, $\alpha$=2.35) we might 
expect a total mass of stars of $\sim$750~$M_{\odot}$  for every star with an  initial mass of  
$\sim$30~$M_{\odot}$, leading to a mass estimate of  a few $\times 10^4M_{\odot}$ (S. Goodwin priv. comm. 2002). 

This {\em lower} limit to the initial mass  of \object{Wd~1} suggests that it is directly  comparable to the galactic centre clusters such as
 the \object{Arches}  cluster (4$\times$10$^4$~$M_{\odot}$; Portegies Zwart et al. \cite{portegies}).  More reasonable estimates of completeness and 
progenitor mass suggest a mass for Wd1 of a few $\times 10^5 M_{\odot}$ making it by far the most massive young Galactic cluster, and 
one of the most massive in the Local Group. This conclusion is further reinforced by 
considering the large number of YHGs within \object{Wd~1},  which is comparable
to the total population of the Milky Way (6; de Jager \cite{dejager}). Assuming a lifetime for the YHGs of  $\sim$25000~yr, 
following the arguments of Geballe et al.  (\cite{geballe}) for the likelihood of finding several examples of a short lived evolutionary 
phase within a single cluster also leads to the conclusion  that a very large O star population ($\sim$several hundred) 
is required to produce the number of YHGs observed. 

Evidently a combined spectroscopic and photometric approach to both 
identify the MS turn-off and to properly classify evolved stars will be 
required to accurately determine the age and mass of Wd1.  However if the 
above estimates are correct then \object{Wd~1} would appear to be a Galactic 
equivalent of the super star clusters observed in merging and interacting 
galaxies and may possibly be more massive than the \object{30~Doradus} cluster in 
the LMC.  Therefore, as well as providing a unique laboratory for 
studying hot star evolution, \object{Wd~1} would provide an unprecidented insight 
into an extreme mode of cluster formation, previously not thought to be 
occuring in the Milky Way.

\begin{table}
\caption{Transitions and equivalent widths for the WN 
candidates. Errors are estimated at 10\% for lines longwards of 
$\sim$7000{\AA} and 20\% for those shortwards. Transitions 
for which the S/N is too poor to attempt an identification 
are indicated with `S/N'.}
\begin{center}
\begin{tabular}{lcccccc}
\hline
Candidate         & A   & G   & I & B & J & D \\
Transition    &         &  &   &     &   &     \\
\hline
He\,{\sc ii}  6560 &27 & S/N & 9 &S/N&S/N& S/N  \\             
He\,{\sc i}  6678+He\,{\sc ii}  6683      & -  & 30  & 37&S/N&S/N& S/N  \\ 
C\,{\sc iv}   7061+He\,{\sc i}   7065    &   -  &  59 &50 & 45& - & 46\\   
N\,{\sc iv} 7103-28& 110 &  Bl & Bl & Bl & 19& Bl \\ 
He\,{\sc ii} 8237   &  33 & 20  & 19& 12&  7& 11\\ 
HeII 5-8? &  38 & -   &  -&  -& -  & -  \\ 
He\,{\sc ii} 10124 & 106 & 56  & 42& 33& 32& 41\\ 
\hline
\end{tabular}
\end{center}
\end{table}

\begin{table}
\caption{Transitions and equivalent widths for the WC candidates; terminology as for Table 2.}
\begin{center}
\begin{tabular}{lccccc}
\hline
Candidate                           & F   & E   & C & H & K \\
Transition    &            &     &   &   &    \\
\hline
He\,{\sc ii} 6560+C\,{\sc ii} 6578  &42               &49   &S/N& S/N & 13   \\
He\,{\sc i} 6678                         &9                &17   &S/N& S/N & -   \\
C\,{\sc ii} 6725-42+C\,{\sc iii} 6727-73&Bl &Bl &S/N& S/N &  -  \\
C\,{\sc ii} 6780                        &28               &45   &S/N& S/N &  -  \\
C\,{\sc iii} 7037                      &Bl &Bl &S/N& S/N &  -  \\
He\,{\sc i} 7065+C\,{\sc ii} 7064   &15               &32   &S/N& S/N &  -  \\
C\,{\sc ii} 7236+C\,{\sc iii} 7210-12 &55               &108  &45 & 24  & 3   \\
C\,{\sc iii} 7772-96                      &S/N              &8    &S/N& -  &  -  \\
C\,{\sc iii} 8196                     &12               &15   &16 & 5  &  -  \\
C\,{\sc iii} 8256                   &Bl   &9    &Bl  & -  &  -  \\
C\,{\sc iii} 8328-59                  &39               &27   &18 & 10  & -   \\
C\,{\sc iii} 8500                  &14               &17   &11 & 6  &  -  \\
C\,{\sc iii} 8664              &19               &12   & - & -  &  -  \\
He\,{\sc ii} 9225+      C\,{\sc ii} 9224  &16               &25   &20 & 14 &  -  \\
C\,{\sc iii} 9705                      &103              &139  &97 & 50  & 26   \\
C\,{\sc ii} 9903                       &18               &25   &S/N & 9  &  -  \\
\hline
\end{tabular}
\end{center}
\end{table}

\begin{table}
\caption{Numbers of B \& RSGs and WRs for galactic open clusters containing 
one or more of each type of object. The numbers of B \& RSGs for \object{Wd~1} 
are those from West87; preliminary analysis of our data suggest they are likely to be 
lower limits. Note that the Quintuplet cluster also contains 5 `Cocoon' objects with 
featureless IR spectra that have been proposed as dusty WCL stars.
($^a$Massey, DeGioia-Eastwood \& Waterhouse
\cite{massey01} and references therein; $^b$Figer, McLean \& Morris M. \cite{figer};
$^c$Paumard et al. \cite{paumard}; $^d$Carr, Sellgren \& Balachandran \cite{carr};
$^e$Blum et al. \cite{blum01}).}
\begin{center}
\begin{tabular}{lcccc}
\hline
Cluster & Age(Myr) & $N_{\rm BSG}$ & $N_{\rm RSG}$ & $N_{\rm WR}$ \\
\hline
{\bf Wd~1}  & 4-8 & $\geq$6 & $\geq$2 &$\geq$11 \\ 
Trumpler 27$^a$ & NA & 8     &    1 & 2 \\
Collinder 228$^a$ & $\sim$2 &3 & 1 & 1 \\
Berkley 87$^a$ & $\sim$3&4 & 1 & 1 \\
Quintuplet$^b$ & 4$\pm$1&14 (+ 2 LBVc) & 1 & 8 \\
Gal. Center$^{c,d}$ & 3-8 & 7 He\,{\sc i}  & 1 & 9 He\,{\sc i} \\
Arches$^e$ & 2-4.5 & 0 & 0 & 15 \\
\hline
\end{tabular}
\end{center}
\end{table}

\section{Acknowledgements}
We thank
S. Goodwin, R. Stothers, R. Waters, S. Dougherty,
 P. Crowther, K. de Jager and  H. Nieuwenhuijzen for
 informative discussions and the referee, W.-R. Hamann, for 
his constructive comments.

\end{document}